\documentclass[useAMS,usenatbib]{mn2e}
\usepackage{journals}
\usepackage{graphicx}
\newcommand{\oversim}[2]{\protect{\mbox{\lower0.5ex\vbox{%
  \baselineskip=0pt\lineskip=0.2ex
  \ialign{$\mathsurround=0pt #1\hfil##\hfil$\crcr#2\crcr\sim\crcr}}}}}
\newcommand{\simgreat}{\mbox{$\,\mathrel{\mathpalette\oversim>}\,$}} % >~ sign
\newcommand{\simless} {\mbox{$\,\mathrel{\mathpalette\oversim<}\,$}} % <~ sign
\title[Top-heavy IGIMFs in starbursts]{Top-heavy
  integrated galactic stellar initial mass functions (IGIMFs) in starbursts}
\author[C.~Weidner, P.~Kroupa,
  J.~Pflamm-Altenburg]{C.~Weidner$^{1,2}$\thanks{E-mail:~cw60@st-andrews.ac.uk},
  P.~Kroupa$^{3}$\thanks{E-mail:~pavel@astro.uni-bonn.de} and
  J.~Pflamm-Altenburg$^3$\thanks{E-mail:~jpflamm@astro.uni-bonn.de}\\
$^{1}$Scottish Universities Physics Alliance (SUPA), School of Physics and
  Astronomy, University of St. Andrews, North Haugh,\\
 St. Andrews, Fife KY16 9SS, UK\\
$^{2}$Departamento de Astronom{\'i}a y Astrof{\'i}sica, Pontificia
 Universidad Cat{\'o}lica de Chile, Av.~Vicu{\~n}a MacKenna 4860, Macul,\\
      Santiago, Chile\\
$^{3}$Argelander-Institut f\"ur Astronomie (Sternwarte), Auf dem H{\"u}gel 71,
D-53121 Bonn, Germany}

\begin{document}
\bibliographystyle{aa}
\date{Accepted . Received ; in original form }

\pagerange{\pageref{firstpage}--\pageref{lastpage}} \pubyear{2010}

\maketitle

\label{firstpage}

\begin{abstract}
Star formation rates (SFR) larger than 1000 $M_\odot$ yr$^{-1}$ are
observed in extreme star bursts. This leads to the formation of star
clusters with masses $>$ 10$^6$ $M_\odot$ in which crowding of the
pre-stellar cores may lead to a change of the stellar initial mass
function (IMF). Indeed, the large mass-to-light ratios of
ultra-compact dwarf galaxies and recent results on globular clusters
suggest the IMF to become top-heavy with increasing star-forming density.
We explore the implications of top-heavy IMFs in these
very massive and compact systems for the integrated galactic
initial mass function (IGIMF), which is the galaxy-wide IMF, in
dependence of the star-formation rate 
of galaxies. The resulting IGIMFs can have slopes, $\alpha_3$, for stars
more massive than about 1 $M_\odot$ between 1.5 and the Salpeter slope
of 2.3 for an embedded cluster mass function (ECMF) slope ($\beta$) of
2.0, but only if the ECMF has no low-mass clusters in galaxies with
major starbursts. Alternatively, $\beta$ would have to decrease with
increasing SFR $>$ 10 $M_\odot$ yr$^{-1}$ such that galaxies with
major starbursts have a top-heavy ECMF. The resulting IGIMFs are 
within the range of observationally deduced IMF variations with
redshift.
\end{abstract}


\section{Introduction: Clustered star-formation and the IMF}
\label{se:intro}
The stellar initial mass function (IMF) is one of the fundamental
astrophysical distribution functions. It defines the ratio of
low-mass stars, which do not contribute to the chemical evolution over a
Hubble time but lock-up baryonic matter, to high-mass stars,
which power the interstellar medium and enrich it with metals through
AGB-winds and supernovae. It further determines the mass-to-light
ratios of stellar populations and influences the dynamical evolution
of star clusters and whole galaxies.

The low-mass end of the IMF is found to be independent
of environmental influences like metallicity and density, which may be
understandable theoretically by a nearly constant Jeans-mass due to
the way how molecular cooling rates scale with density and temperature
\citep{EKW08}. In the regime around and below the hydrogen burning
mass-limit it now seems that brown dwarfs form an individual
distribution disjoint but related to the low-mass IMF
\citep{TK07,TK08}. The formation and theoretical basis of the IMF of
massive stars ($>$ 10 $M_\odot$) is less well understood with at least
two competing theories (competitive accretion vs.~single star
accretion) having been developed
\citep{BBZ98,BVB04,BB06,TKM06,KKM09}. Observationally, though, the 
slope of the high-mass IMF within star clusters seems to be as
independent from the environment as the low-mass slope. Unresolved
multiple systems have been shown to have no effect on the $m >$ 10
$M_\odot$ IMF \citep{MA08,WK07c} but for $m \le$ 1 $M_\odot$ unresolved multiple
stars effect the observed slope of the IMF substantially \citep{KGT91}.

Over the last years it has become clear that star formation takes place
mostly in embedded clusters \citep{LL03}, each cluster containing a
dozen to many million of stars \citep{Kr04b}. Within these clusters stars
appear to form following the canonical IMF, $\xi(m) \propto m^{-\alpha}$,
with a slope of 1.3 for low mass stars and the Salpeter/Massey-slope
of 2.35 for massive stars (for more details on the canonical IMF see
Appendix~\ref{app:IMF}).

Not only stars follow a mass function but also (young, embedded) star
clusters. The embedded cluster mass function (ECMF) has been found to
be a power-law, $\xi_\mathrm{ecl} \propto M_\mathrm{ecl}^{-\beta}$, with
a rather constant slope of $\approx 2$ for largely different
environments from the quiescent solar neighbourhood to the vigorously
star-forming Antennae galaxies \citep{LL03,HEDM03,ZF99}.

The upper mass end of the ECMF (the most-massive young cluster within a
star-forming galaxy) has been found to depend on the SFR of the
galaxy \citep{WKL04,Ba08}. Furthermore, it appears that
that the mass of the most-massive star in a
star cluster is related to the mass of the cluster in a non-trivial
way \citep{WKB09}. Low-mass clusters seem to be unable to form very massive
stars. The physical reason for this empirical relation has not yet
been found. But it can be presumed that the interplay between
stellar feedback (ionising radiation and winds) from the massive stars
and the gravitational potential of the star-forming cloud might play a
role. In this picture, the feedback of the massive stars
overcomes the binding energy of the cloud and star-formation is
terminated. Such a process would directly couple the mass of the cloud
with the feedback of the stars which is directly correlated with their
mass \citep{El83}. The majority of these embedded star clusters then dissolve
quickly due to gas-expulsion \citep{KAH01,AM01,KB02,LL03,FMFM04,PGK09}.

A direct consequence from clustered star-formation is that the
composite stellar population in a galaxy, which results from many
star-forming events, is the sum of the dissolving star clusters. Thus
the integrated galactic initial mass function (IGIMF) is
the sum of all the IMFs of all the star clusters \citep[][see also \citealt{VB82}]{KW03,WK05a},

\begin{eqnarray}
\label{eq:IGIMF}
\xi_{\rm IGIMF}(m, t) &=& \int_{M_{\rm ecl,min}}^{M_{\rm ecl,max}(SFR(t))} \xi(m
\le
                      m_{\rm max}(M_{\rm ecl})) \nonumber \\
& &\cdot\,\xi_{\rm ecl}(M_{\rm ecl})\,dM_{\rm ecl},
\end{eqnarray}

where $M_{\rm ecl,min}$ is the minimal embedded cluster mass, 
$M_{\rm ecl,max}(SFR(t))$ is the maximum embedded cluster mass which
is dependent on the SFR of the galaxy and is given by the following
equation \citep{WKL04},
\begin{equation}
\label{eq:eclmax}
M_\mathrm{ecl,max} = 8.5 10^{4} \times SFR^{0.75}~M_\odot.
\end{equation}

And $\xi_{\rm ecl}(M_{\rm ecl})$ is the ECMF\footnote{Note that, strictly,
$\xi_\mathrm{ecl}$ describes the distribution of star-forming
molecular cloud cores containing only the stellar mass formed. The
``embedded clusters'' in eq.~\ref{eq:IGIMF} do not, under any
circumstances, mean bound or radially well-defined stellar
ensembles. Rather, eq.~\ref{eq:IGIMF} is an integral over all
locally correlated star forming events, a small fraction of which
will hatch from the clouds as bound clusters, while the majority
disperse within about 10 Myr.}, with $dN_\mathrm{ecl} = \xi_{\rm
ecl}(M_\mathrm{ecl})\,dM_\mathrm{ecl}\,\propto\,M_\mathrm{ecl}^{-\beta}
dM_\mathrm{ecl}$ being the number of just formed embedded clusters
with stellar mass in the interval $M_\mathrm{ecl}$, $M_\mathrm{ecl} +
dM_\mathrm{ecl}$. Finally, $\xi(m\le m_{\rm max}(M_{\rm ecl}))$ is
the IMF within an embedded star cluster with the total stellar mass
$M_\mathrm{ecl}$, with the most-massive star , $m_\mathrm{max}$,
which is dependent on $M_\mathrm{ecl}$ \citep{WK05b,WKB09}. Note that
eq.~\ref{eq:eclmax} is originally \citep{WKL04} only fitted to SFRs
of up to 4 $M_\odot$ yr$^{-1}$ and extra-polated for starbursts with
SFRs $>$ 10  $M_\odot$ yr$^{-1}$ in this work. But it has been
recently shown \citep{Ba08} that eq.~\ref{eq:eclmax} also holds for higher
SFRs.

The IGIMF is steeper than the individual canonical IMFs in the
actual clusters, hereby immediately explaining why $\alpha_{3,
  \mathrm{field}} = 2.7 > \alpha_3 = 2.35$, where
$\alpha_{3,\mathrm{field}}$ is the slope of the IMF derived by
\citet{Sc86} and \citet{RGH02} from OB star counts in the Milky Way
field, and $\alpha_3$ = 2.35 is the \citet{Sal55} index. This is due
to the fact that low-mass star clusters are numerous but can not have
massive stars.

The high mass part of the resulting IGIMF is strongly dependent on the
SFR of a galaxy \citep[][]{HA10}. This result has 
been found to be  naturally (without parameter adjustments) able to
explain the mass-metallicity relation of galaxies \citep{KWK05}, the
alpha-element abundances as a function of galaxy mass \citep{RCK09}
and the H$\alpha$ cut-off in star-forming galaxies \citep{PAK08}. It
also predicted a discrepancy between SFRs derived from UV- and
H$\alpha$-fluxes \citep{PWK07,PWK09}, a result which has recently
been confirmed qualitatively \citep{MWK09} and quantitatively
  \citep{LGT09} by observations. Furthermore, the
H$\alpha$-SFR relation calculated in the IGIMF theory leads to higher
SFRs of H$\alpha$ faint star-forming dwarf galaxies \citep{PWK07} and
reveals a simple linear relation between the total neutral gas mass
and SFRs of galaxies \citep{PK09}.

The IGIMF theory (eq.~\ref{eq:IGIMF}) is based solely on observed
correlations ($m_\mathrm{max}$ vs $M_\mathrm{ecl}$,
$M_\mathrm{ecl,max}$ vs $SFR$) and distribution functions (IMF in the
correlated star formation events, or ``embedded clusters'', and the
mass function of stellar masses formed in these events, the ECMF). It
is these observed correlations and distribution functions which
contain the relevant star-formation physics. It is this basis which
allowed the IGIMF theory to be so successful since its first
formulation \citep{KW03}. The empirical basis of the IGIMF
theory is important because star-formation theory is not advanced
enough to make reliable statements on the galaxy-wide IMF, and would
in any case lead directly to the IGIMF theory since star-formation
theory will have to account for the observed correlations and
distribution functions.

In the case of very high SFRs ($>$ 30 $M_\odot$/yr) the
$M_{\rm ecl,max}-SFR$-relation results in very massive star clusters
($M_\mathrm{ecl} > 10^6 M_\odot$) which may be the progenitors
of the Ultra Compact Dwarf galaxies (UCD), as
these most-likely formed as compact objects of a few pc size
\citep{DKB09}. And indeed recent HST/Spitzer observations found very
young massive ($10^9 M_\odot$) objects with radii below 100 pc in
low-redshift Lyman-Break-Analog galaxies with high
star-formation rates \citep{OHT09}.

In this contribution we explore the consequences of very high
galaxy-wide SFRs on the IGIMF by taking into account recently acquired
evidence for top-heavy IMFs when UCDs and possibly also globular
cluster are formed. This advances the IGIMF theory into a physical
regime of cosmological significance since very high SFRs are typically
observed at high redshifts \citep{TIZ07}.

The observational situation regarding top-heavy IMFs and crowding in
massive proto-clusters is discussed in Section~\ref{se:obs}, while the
top-heavy IGIMF model is presented in Section~\ref{se:model}. The following
Section~\ref{se:res} then shows the results of the model
calculations which are discussed in Section~\ref{se:disc}. The canonical IMF
within star clusters used throughout this work is presented in
Appendix~\ref{app:IMF}.

\section{Observational Situation}
\label{se:obs}
\subsection{Top-heavy IMF}
\label{sub:top}
Despite all the evidence for a universal IMF, recently several
indications have begun emerging for a possible dependence of the shape
of the IMF on environment. Several observational and theoretical
indications suggest the IMF to become top-heavy under extreme
starburst conditions. For example, \citet{BLF05} find that a flatter high-mass
slope of the IMF at higher redshifts is needed to explain the observed
numbers of sub-mm galaxies. \citet{SMB05} observe rather steep slopes in the
spectral energy distribution of high-redshift objects in the Hubble
Ultra Deep Field. They conclude this can be either explained by a 50\%
higher star formation rate than the current calibrations yield or a
top-heavy IMF at very high-redshifts (e.g.~$z\,>$ 10-15). This is
actually predicted by \citet{CB03}. A similar evolution of the IMF with
redshift such that at high-$z$ star-formation is biased towards more
massive stars is found by studying the amplitude of the
galaxy-stellar-mass--star-formation-rate relationship \citep{Da07} and
by comparing the rate of the luminosity evolution of massive
early-type galaxies in clusters to the rate of their colour evolution
\citep{VD07}. As star-formation rates are also higher at high-$z$ this
can be seen as an indication for a top-heavy IMF in starbursts. When
integrating the redshift evolution of the SFR over cosmological
time-scales within the $\Lambda$ cold dark matter cosmology
\citet{WHT08} find that they need a top-heavy IMF with $\alpha_3$ =
2.15 to reproduce the present-day stellar mass density. However, in
order to explain the stellar mass density at higher redshifts an even
flatter slope is needed. Other constrains also come from the element
abundances observed in galaxies. \citet{FBB03} study the element
abundances in groups and clusters of galaxies with the use of {\it
XMM-Newton} and {\it ASCA}. They show ``{\it that while the metal
production in groups could be described by a stellar population with a
standard local initial mass function, clusters of galaxies require a
more top-heavy IMF.}`` From modelling the iron evolution in galaxy 
clusters and the field and comparing the results with observations,
\citet{Lo06} concludes that a top-heavy IMF is indeed needed in galaxy
clusters but not for galaxies in the field. \citet{TBM04} find a
similar result in their $N$body-Tree + SPH simulations of a galaxy
cluster. A Salpeter IMF produces subsolar [O/Fe] abundances. They can
explain solar [O/Fe] if the IMF is top-heavy for $z\,>$ 2 and Salpeter
afterwards. To account for the [$\alpha$/Fe]-$\sigma$-relation observed in local elliptical galaxies 
\citet{CM09} also require an top-heavy IMF. \citet{BKM07,BMO07} infer the need of a top-heavy IMF in
the Galactic bulge in order to 
explain the difference in the [O/Mg] to [Mg/H] ratios in bulge Giants
compared to the Galactic disk. But \citet{MMB07} find that
metallicity-dependent stellar yields for massive stars equally explain
the ratios. Hints for a top-heavy IMF are also found in the centre of
the Milky Way. By examining the evolution of disk stars orbiting
a central black hole \citet{ABA07} found strong evidence for a
top-heavy IMF in order to explain the observed ring of massive
stars orbiting about 0.1pc around the Galactic centre. Using {\it Chandra}
X-ray data of the Galactic centre \citet{NS05} found 10 times less
X-ray emission from low-mass stars than expected if the observed massive
stars formed from the canonical IMF. \citet{El04b} did an extensive
literature study of Galactic and extragalactic observations and
concluded that dense star-forming regions like starbursts might have a
slightly shallower IMF, a view shared by
\citet{Ei01b,Ei01}. Furthermore, \citet{BGK07} find that 
the majority of observed hypervelocity stars ejected by the
super-massive black hole in the centre of the Galaxy have a mass
around 3 or 4 $M_{\odot}$ whilst lower mass stars are lacking from
observations. The high number of carbon-enhanced,
s-process-enriched unevolved stars among extremely metal-poor stars in
the halo of the MW has been interpreted by \citet{LGB05} as evidence
for a top-heavy IMF during the formation of the halo. A recent study
found that the central regions of disturbed galaxies host far more
supernovae Ib and Ic (originating from very massive stars) while in
all other environments supernova typ II dominate \citep{HAJ10}. This
can be attributed to a top-heavy IMF in starbursts after galaxy
interactions and mergers.

Marks, Kroupa \& Dabringhausen (in preparation) analyse the stellar
content of globular clusters and the dynamical M/L ratios of UCDs
using $N$-body models and deduce that the 
IMF appears to become top-heavy with increasing density of the forming
object. Given the mass-radius relation inferred from the
proto-globular clusters and proto-UCDs this translates into a
dependence of the IMF slope $\alpha_3$ on the object's mass. The
application of the IGIMF theory to high galaxy-wide SFRs is based on
this result. \citet{Pa10} shows that the molecular gas is likely
to have a higher temperature in massive star bursts as a result of heating
through supernova generated cosmic rays. This may be a theoretical
reason for expecting IMFs that are top-heavy in such systems.

\subsection{Crowding in very massive pre-cluster cloud cores}
\label{sub:crowding}

A substantial fraction of stars seems to form in star cluster-like
correlated events \citep{AM01,LL03} and these are observed to be
very compact during their final stages of collapse \citep[$r_{\rm 
    ecl}\,\simless\,1$~pc,][]{TPN98,Kr04b,GMP05,TLK05,RJS06,SGH06} and
are therefore in very dense configurations.

Also, strikingly similar values are found for the size of pre-stellar cloud
cores, despite large differences in mass. \citet{FKS06} find radii
of about 5000 AU for a low-mass star-forming environment,
\citet{BAW01} find 5000 to 10000 AU and \citet{vdT00} find 20000 AU
for massive star progenitors.

Furthermore, within starbursts extreme star formation rates
\citep[$>$ 100 $M_{\odot}/{\rm yr}^{-1}$,][]{YCF04,DCR04,CYI06}
are reached, e.q.~in the case of Arp 220 \citep{WHL06}. The recent
Herschel study by \citet{MLB10} of submillimetre galaxies (SMG) 
found that for SMGs with fluxes $S(850\mu\mathrm{m})$ $>$ 5 mJy the
median SFR is 960 $M_\odot$ yr$^{-1}$. The highest
$S(850\mu\mathrm{m})$ flux in their sample is about 10 mJy,
corresponding to a SFR of several $10^3\,M_\odot$ yr$^{-1}$.

This leads to the formation of very massive star clusters
according to eq.~\ref{eq:eclmax}. For the characteristic time-scale
of conversion of the interstellar medium to stars, $\delta t$, of about
10 Myr \citep{ESN04,EKS09}, the total mass in stars is
$M_\mathrm{tot}$ = SFR $\times$ $\delta t$. But as also
$M_\mathrm{tot} = \int_{M_\mathrm{ecl,min}}^{M_\mathrm{ecl,max}} M_\mathrm{ecl}
\xi_\mathrm{ecl}(M_\mathrm{ecl})dM_\mathrm{ecl}$, $M_\mathrm{ecl,max}$
increases as well when the SFR increases. The normalisation of
$\xi_\mathrm{ecl}(M_\mathrm{ecl})$ is obtained by stating there is one
most-massive cluster, 1 = $\int_{M_\mathrm{ecl,max}}^{\infty}
\xi_\mathrm{ecl}(M_\mathrm{ecl}) dM_\mathrm{ecl}$, forming during each
10 Myr interval.

Because young pre-star clusters are compact ($r \approx 1$ pc) and
since the contracting pre-stellar cores within them have dimensions of
a few thousand AU the pre-stellar cores are likely to begin
interacting and coalescing above a certain mass limit. \citet{BBZ98}
predict an influence on the IMF of the forming star cluster, and so do
\citet{ES03} who investigated this question of crowding further. They
concluded that crowding should be relevant in the most massive
clusters (e.g.~progenitors of globular clusters) leading to a
top-heavy IMF, because the crowded medium would have a 
higher low-mass limit for star formation and therefore more massive
stars need to be build to result in the same total mass. \citet{El04a}
constructed a three-component IMF model from these
results. \citet{Sh04} further showed that the effect of crowding must have
an influence on the IMF in starburst clusters. While \citet{El04a} and
\citet{Sh04} present models for the IMF in a starburst they do not
discuss the overall integrated IMF for a whole galaxy with a
starburst. Certainly, in massive clusters, or UCDs, the IMF will be
effected by crowding \citep[higher M/L ratio, ][]{DKB09}, but the
question remains if this change is large enough to influence the
IGIMF significantly. 

\begin{figure}
\begin{center}
\includegraphics[width=8cm]{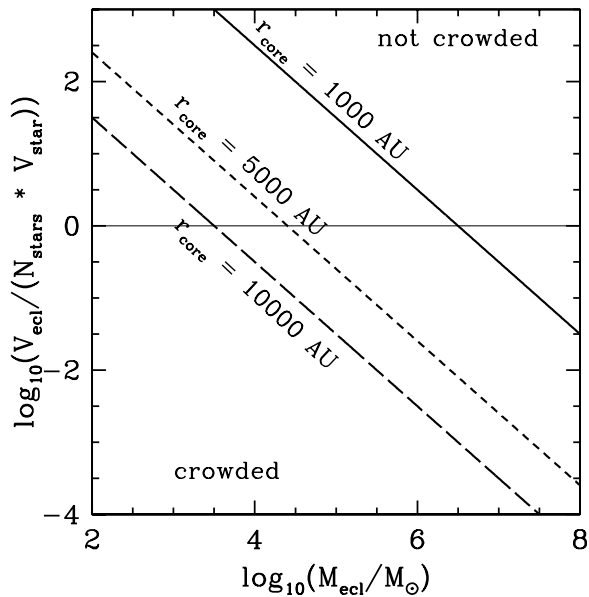}
\vspace*{-1.5cm}
\caption{The volume of an embedded star cluster ($V_{\rm ecl}$, with a
  radius of 1 pc) divided by the volume of all pre-stellar cores
  within it ($N_{stars}~*~V_{\rm star}$) versus the embedded cluster mass
  ($M_{\rm ecl}$). The {\it thick solid line} indicates the assumption
  of a pre-stellar core radius of 1000 AU, while the {\it thick
  short-dashed line} assumes a pre-stellar core radius of 5000 AU and
  in the case of the {\it thick long-dashed line}, a pre-stellar core
  radius of 10000 AU is used. The value when the volume of all stars
  equals the volume of the cluster is indicated by a {\it horizontal
    solid line}.}
\label{fig:crowding}
\end{center}
\end{figure}

In Fig.~\ref{fig:crowding} the crowding is illustrated. It shows how
many proto-stars of a fixed pre-stellar cloud core size can be fitted
into a spherical proto-cluster with a radius of 1 pc. This is done by
taking a range of proto-cluster masses and dividing these masses by the
mean mass of the canonical IMF ($m_\mathrm{mean} \approx$ 0.36
$M_\odot$) in order to get the expected number of stars,
$N_\mathrm{expec}$, in the cluster. $N_\mathrm{expec}$ is then
multiplied by the volume of three different pre-stellar cloud core
sizes - 1000 AU ({\it thick solid line}), 5000 AU ({\it thick
short-dashed line}), and 10000 AU ({\it thick long-dashed
line}). Fig.~\ref{fig:crowding} shows the volume of the proto-cluster
divided by the total volume of the pre-stellar cloud cores. For
clusters with more than $\approx 10^{5}\,M_{\odot}$ crowding will be a
problem. It is therefore appropriate to assume that such high-density
environments may have an influence on the IMF within such star
clusters.

An alternative or additional mechanism which may induce star bursts
leading to top-heavy IMFs may be through the heating of the molecular
gas by supernova generated cosmic rays \citep{Pa10}.

\section{The Model}
\label{se:model}
While the above discussion suggests that the IMF may become top-heavy,
it remains, unfortunately, unclear as to how a systematic variation of
the IMF with increasing $M_\mathrm{ecl}$ ought to be realised.
\citet{El04a} and \citet{Sh04} already presented descriptions of
the change of an IMF in starbursts. They based their models on a
log-normal IMF \citep{Sc86} but we use the multi-power law formulation 
of the stellar IMF.

The modelling of the IGIMF starts from a SFR which determines the upper
mass limit of the ECMF according to eq.~\ref{eq:eclmax}. In order to
calculate the total amount of stars formed in a star-forming period
the SFR is multiplied by a time-step which is assumed to be 10 Myr.
With the use of the total mass the ECMF can be
normalised for the time-step. The ECMF is then divided in 1000
logarithmic bins. As the embedded cluster mass limits the most-massive
star in each cluster, the upper limit of the IMF in each ECMF-bin can
be calculated. As the number of clusters per bin is given by the
normalised ECMF, the IMFs of each ECMF-bin can be multiplied by the
number of clusters in order to give the integrated galactic IMF of the 
time-step. The slope of this IGIMF above 1.3 $M_\odot$ can then be
computed by a least-squares fit to the calculated IGIMF. The value of
the so-derived $\alpha_3$ gives a rough indication of the
top-heaviness, while detailed astrophysical parameters would be
derived from using the actually computed IGIMF.

In order to include the possible effect of crowding into the framework
of the IGIMF theory (\S~\ref{sub:crowding}) the results of Marks,
Kroupa \& Dabringhausen (in preparation) on the high-mass IMF slope in
globular clusters and UCDs are used. They derive the following
dependence of the IMF slope for stars more massive than 1 $M_\odot$
and for clusters with initial masses $M_\mathrm{ecl}$ $\ge$ 2$\times
10^5$ $M_\odot$,
\begin{equation}
\label{eq:a3M}
\alpha_3(M_\mathrm{ecl}) = -1.67 \times
\log_{10}\left(\frac{M_\mathrm{ecl}}{10^6 M_\odot}\right) + 1.05.
\end{equation}
The limit of 2$\times 10^5$ $M_\odot$ is chosen because for
clusters with masses below this limit Marks, Kroupa \& Dabringhausen
(in preparation) results infer a Salpeter slope. Note that while
  the parametrisation in eq.~\ref{eq:a3M} rests on empirical globular
  cluster data which have been corrected for stellar and dynamical
  evolution, the parametrisation we adopt in eq.~\ref{eq:a3M} is not
  to be seen as an established dependence of $\alpha_3$ on
  $M_\mathrm{ecl}$. Rather, while giving a good clue as to how the
  IGIMF may change with increasing SFR, the calculations performed
  here are independent of the parametrisation and may thus be easily
  adopted for different parametrisations.

\subsection{Model parameters}
\label{sub:SFR}
Besides the change in the IMF for very massive clusters, the most
important parameters in our model are the SFRs of the models (which
give the upper mass limit of the forming clusters,
$M_\mathrm{ecl,max}$), the slope ($\beta$) of the ECMF, and the lower
mass limit of the ECMF, $M_\mathrm{ecl,min}$. For this study of
starbursts four SFRs (10, 100, 1000 and 10000 $M_\odot$ yr$^{-1}$) are
chosen. Below 10 $M_\odot$ yr$^{-1}$ little impact on the IGIMF from the
top-heavy IMFs in very-massive clusters is to be expected as only very
few or none of them are formed according to eq.~\ref{eq:eclmax}. 
As already discussed in the introduction, SFRs of more than 1000
$M_\odot$ yr$^{-1}$ have recently been found \citep{MLB10}. The upper
limit for the SFRs in the models is therefore 
chosen to be 10000 $M_\odot$ yr$^{-1}$. The value of the slope of the
ECMF is usually given with a $\beta$ around 2 with errors between 0.2
and 0.5 \citep{Larsen2009}, though some studies find systematically
flatter slopes like 1.8 \citep{DBT08}. And the mass spectrum of giant
molecular clouds, the precursors of star clusters, shows a slope of
1.7 \citep{R05}. As the identification of embedded clusters in
extra-galactic objects is very challenging, we allow a greater range
for this parameter and study $\beta$ = 0.5 to 2.35. The lower mass
limit of the ECMF, $M_\mathrm{ecl,min}$, is even worse constrained
than the slope of the ECMF. We therefore study a broad range of
possible $M_\mathrm{ecl,min}$ from 5 to 10$^5$ $M_\odot$.

The lower mass limit of the canonical IMF in each cluster is fixed to 0.01
$M_\odot$ and the upper mass limit is determined by the cluster mass
\citep{WK04}.

\section{Results}
\label{se:res}

Tab.~\ref{tab:first} shows the IGIMF slopes for stars more massive than
1.3 $M_\odot$ for different star-formation rates, different values of
the ECMF slope, $\beta$, and for several lower limits of the ECMF,
$M_\mathrm{ecl, min}$. For clusters above 2$\times 10^5$ $M_\odot$ the
IMF slope for stars above 1 $M_\odot$, $\alpha_3$, is changed according
to eq.~\ref{eq:a3M} while below 2$\times 10^5$ $M_\odot$ $\alpha_3$ is
kept constant at the canonical (Salpeter) index 2.35. For $\alpha_3$ a
lower limit of 1 is used because clusters with an $\alpha_3$ of 1 or
less contain more than 90\% of the mass in stars more massive than 8
$M_\odot$. When these stars explode as supernovae the resulting
extreme mass-loss would completely disperse the star clusters. In such
a case no globular clusters would survive until today. For clusters
more massive than $\approx$ 1.1$\times 10^6$ $M_\odot$ this limit is
reached and $\alpha_3$ is fixed to 1.0 for these objects.

\begin{table*}
\caption{\label{tab:first} The IGIMF power-law slopes for a series of
  models with a changing $\alpha_3$ for clusters above $M_{\rm
  limit}$ = 2$\times 10^5$ $M_\odot$ and for different lower limits of
  the ECMF, $M_{\rm ecl,min}$. The upper mass limit of the ECMF is
  chosen according to eq.~\ref{eq:eclmax}. The slopes are obtained by a
  least-squares power-law fit to the IGIMF above 1.3 $M_{\odot}$. The
  columns designated as ``dev'' show the deviation of the top-heavy
  IGIMF from a pure power-law IMF. This is done by dividing the
  number of stars above 8 $M_\odot$ from the top-heavy IGIMF
  model by the number for a pure power-law IMF with the fitted
  slope. Additionally, the columns entitled ``ratio'' list the
  ratio between the number of stars above 8 $M_\odot$ divided by the total
  mass in stars. In the penultimate row, the mean SFR over a Hubble
  time (in $M_{\odot}\,{\rm yr}^{-1}$) is given, which would be needed
  to produce the total stellar mass over 14 Gyr instead of a burst of
  10 Myr. In the final row the maximal cluster mass according
  to eq.~\ref{eq:eclmax} is shown, as well as the ratios
  between the number of stars above 8 $M_\odot$ and the total mass in
  stars for $\beta$ = 2 and $M_{\rm ecl,min}$ = 5 $M_{\odot}$ but
  without a top-heavy IMF in massive star clusters.}
\begin{tabular}{ccccccccccccc}
&slope&slope&slope&slope&dev&dev&dev&dev&ratio&ratio&ratio&ratio\\
\hline
$M_{\rm ecl,min}$&SFR: 10&100&1000&10000&10&100&1000&10000&10&100&1000&10000\\
%$[M_{\odot}]$&$[M_{\odot} {\rm yr}^{-1}]$&$[M_{\odot} {\rm yr}^{-1}]$&$[M_{\odot} {\rm yr}^{-1}]$&$[M_{\odot} {\rm yr}^{-1}]$&$[M_{\odot} {\rm yr}^{-1}]$&$[M_{\odot} {\rm yr}^{-1}]$&$[M_{\odot}{\rm yr}^{-1}]$&$[M_{\odot} {\rm yr}^{-1}]$\\
$[M_{\odot}]$&\multicolumn{4}{c}{$[M_{\odot}\,{\rm yr}^{-1}]$}&\multicolumn{4}{c}{$[M_{\odot}\,{\rm yr}^{-1}]$}&\multicolumn{4}{c}{$[M_{\odot}\,{\rm yr}^{-1}]$}\\
\hline
&\multicolumn{4}{c}{$\beta$ = 0.5}&\multicolumn{4}{c}{$\beta$ = 0.5}&\multicolumn{4}{c}{$\beta$ = 0.5}\\
\hline
5      &1.96  &1.22  &1.03  &1.00&1.00&0.93&0.99&1.01&0.016&0.019&0.019&0.019\\
100    &1.96  &1.22  &1.03  &1.00&1.00&0.93&0.99&1.01&0.016&0.019&0.019&0.019\\
1000   &1.96  &1.22  &1.03  &1.00&1.00&0.93&0.99&1.01&0.016&0.019&0.019&0.019\\
10000  &1.96  &1.22  &1.03  &1.00&1.00&0.93&0.99&1.01&0.016&0.019&0.019&0.019\\
100000 &1.92  &1.21  &1.02  &1.00&1.00&0.94&1.00&1.01&0.017&0.019&0.019&0.019\\
\hline
&\multicolumn{4}{c}{$\beta$ = 1.0}&\multicolumn{4}{c}{$\beta$ = 1.0}&\multicolumn{4}{c}{$\beta$ = 1.0}\\
\hline
5      &2.04  &1.39  &1.12  &1.03&1.00&0.88&0.96&0.99&0.015&0.019&0.019&0.019\\
100    &2.04  &1.39  &1.12  &1.03&1.00&0.88&0.96&0.99&0.015&0.019&0.019&0.019\\
1000   &2.04  &1.39  &1.12  &1.03&1.00&0.88&0.96&0.99&0.015&0.019&0.019&0.019\\
10000  &2.03  &1.39  &1.11  &1.03&1.00&0.88&0.96&0.99&0.015&0.019&0.019&0.019\\
100000 &1.97  &1.34  &1.09  &1.02&1.00&0.91&0.97&1.00&0.016&0.019&0.019&0.019\\
\hline
&\multicolumn{4}{c}{$\beta$ = 1.6}&\multicolumn{4}{c}{$\beta$ = 1.6}&\multicolumn{4}{c}{$\beta$ = 1.6}\\
\hline
5      &2.21  &1.75  &1.50  &1.33&1.00&0.81&0.82&0.87&0.012&0.016&0.018&0.019\\
100    &2.20  &1.74  &1.49  &1.33&1.00&0.81&0.83&0.87&0.012&0.016&0.018&0.019\\
1000   &2.19  &1.73  &1.48  &1.31&1.00&0.82&0.83&0.88&0.012&0.016&0.018&0.019\\
10000  &2.15  &1.68  &1.44  &1.28&1.00&0.82&0.84&0.89&0.013&0.017&0.018&0.019\\
100000 &2.03  &1.54  &1.32  &1.19&1.00&0.86&0.89&0.93&0.015&0.018&0.019&0.019\\
\hline
&\multicolumn{4}{c}{$\beta$ = 2.0}&\multicolumn{4}{c}{$\beta$ = 2.0}&\multicolumn{4}{c}{$\beta$ = 2.0}\\
\hline
5      &2.52  &2.32  &2.16  &2.05&1.00&0.88&0.78&0.74&0.007&0.009&0.010&0.011\\
100    &2.36  &2.15  &2.01  &1.91&1.11&0.88&0.81&0.77&0.010&0.011&0.012&0.013\\
1000   &2.29  &2.06  &1.91  &1.81&1.00&0.86&0.80&0.78&0.011&0.013&0.014&0.015\\
10000  &2.24  &1.93  &1.78  &1.68&1.00&0.83&0.80&0.79&0.011&0.014&0.015&0.016\\
100000 &2.08  &1.69  &1.53  &1.45&1.00&0.85&0.84&0.85&0.014&0.017&0.018&0.018\\
\hline
&\multicolumn{4}{c}{$\beta$ = 2.35}&\multicolumn{4}{c}{$\beta$ = 2.35}&\multicolumn{4}{c}{$\beta$ = 2.35}\\
\hline
5      &3.06  &3.04  &3.04  &3.03&1.50&1.12&1.15&1.15&0.003&0.003&0.003&0.003\\
100    &2.47  &2.44  &2.42  &2.41&1.00&1.00&0.98&0.97&0.008&0.008&0.008&0.008\\
1000   &2.36  &2.29  &2.26  &2.24&1.11&0.99&0.97&0.95&0.010&0.010&0.011&0.011\\
10000  &2.29  &2.14  &2.07  &2.04&1.00&0.90&0.85&0.84&0.010&0.012&0.012&0.013\\
100000 &2.12  &1.81  &1.72  &1.68&1.00&0.85&0.83&0.83&0.013&0.016&0.017&0.017\\
\hline
mean SFR:&7.1$\cdot10^{-3}$&7.1$\cdot10^{-2}$&7.1$\cdot10^{-1}$&7.1&&&&&&&&\\
\hline
$M_{\rm ecl,max}$ $[M_\odot]$&4.75$\cdot 10^5$&2.65$\cdot 10^6$&1.48$\cdot
10^7$&8.25$\cdot 10^7$&&&&&0.0064&0.0068&0.0071&0.0073\\
\hline
\end{tabular}
\end{table*}

\begin{figure}
\begin{center}
\includegraphics[width=8cm]{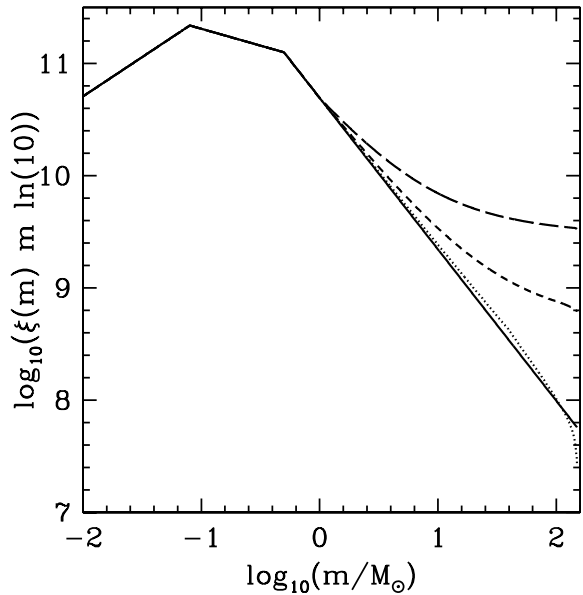}
\vspace*{-1.5cm}
\caption{Three different IGIMFs from Tab.~\ref{tab:first} together
  with the input canonical IMF ({\it solid line}, eq.~\ref{eq:imf}). 
  All three models have $\beta$ = 2.0 and for the
  {\it dotted line} $SFR$ = 10 $M_\odot \mathrm{yr}^{-1}$ and
  $M_\mathrm{ecl,min}$ = 1000 $M_\odot$, for the {\it short-dashed
    line} $SFR$ = 100 $M_\odot \mathrm{yr}^{-1}$, $M_\mathrm{ecl,
    min}$ = 10000 $M_\odot$ and for the {\it long-dashed line} $SFR$ = 1000
  $M_\odot \mathrm{yr}^{-1}$, $M_\mathrm{ecl, min}$ = 100000
  $M_\odot$. All IMFs are normalised to the same total
  mass in stars ($10^{11}\,M_{\odot}$).}
\label{fig:D1F1}
\end{center}
\end{figure}

In Fig.~\ref{fig:D1F1} three IGIMF examples from Tab.~\ref{tab:first}
are shown together with the input canonical IMF. Depending on the 
parameters a strong top-heaviness can be seen.

\begin{figure}
\begin{center}
\includegraphics[width=8cm]{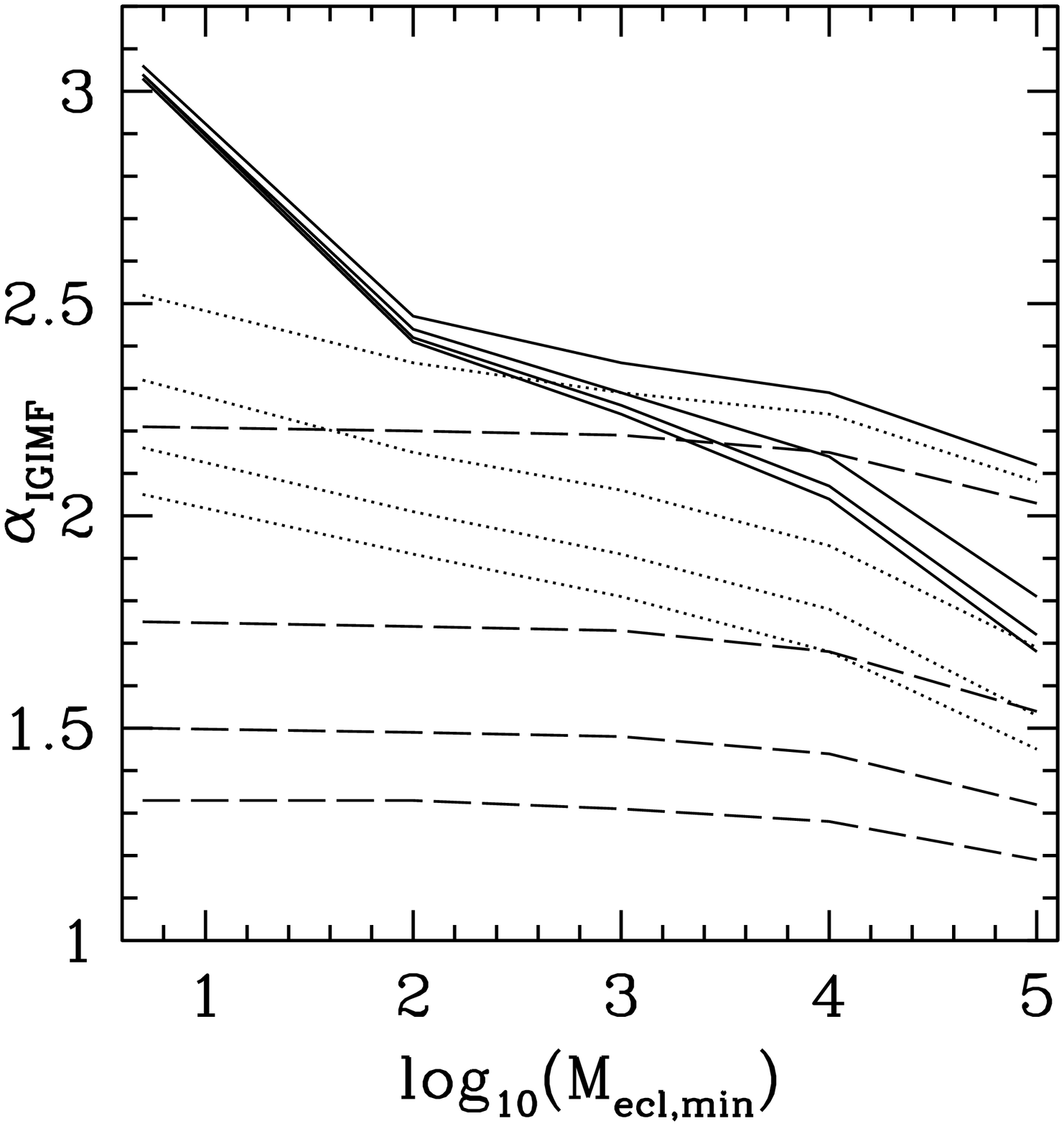}
\vspace*{-1.5cm}
\caption{Dependence of the IGIMF slope above 1.3 $M_\odot$
  ($\alpha_\mathrm{IGIMF}$) on the lower mass limit of the ECMF
  ($M_\mathrm{ecl,min}$). {\it Solid lines} show the results for
  $\beta$ = 2.35 for SFRs from 10 (upper-most line) over 100 and 1000
  to 10000 $M_\odot$ 
  yr$^{-1}$ (lowest line). The {\it dotted lines} are like the {\it
  solid} ones, but for $\beta$ = 2.0 and the {\it dashed lines} for
  $\beta$ = 1.6.}
\label{fig:avsm}
\end{center}
\end{figure}

In Fig.~\ref{fig:avsm} the results of Tab.~\ref{tab:first} are 
summarised for $\beta$ = 2.35, 2.0 and 1.6. Generally, the IGIMF slope
decreases with increasing SFR for a given $\beta$. This is due to the
relation between the maximal cluster mass and the SFR
(eq.~\ref{eq:eclmax}). The higher the SFR the more-massive is the upper
limit of the ECMF. As massive clusters develop a top-heavy IMF
according to eq.~\ref{eq:a3M}, a larger fraction of massive stars are
formed with increasing SFRs. Fig.~\ref{fig:avsm} shows the dependence
of $\alpha_\mathrm{IGIMF}$ on the lower limit of the ECMF
($M_\mathrm{ecl,min}$). For very flat ECMFs ({\it dashed lines}) this
dependency is nearly negligible. This is because for slops of $\beta$
less than 2.0 the ECMF is dominated by massive clusters and reducing
the number of low-mass cluster by increasing $M_\mathrm{ecl,min}$ does
not affect the IGIMF significantly. For $\beta$ = 2.35 the opposite is
the case. For steep slopes, low-mass clusters dominate the ECMF and
therefore have a strong influence on the IGIMF. Here, increasing
$M_\mathrm{ecl,min}$ changes the IGIMF strongly by eliminating the
low-mass clusters.

\section{Discussion}
\label{se:disc}
\begin{figure}
\begin{center}
\includegraphics[width=8cm]{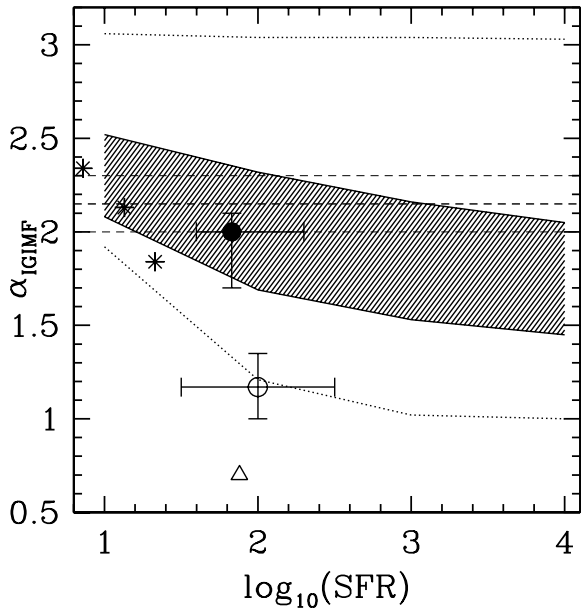}
\vspace*{-1.5cm}
\caption{IGIMF slopes above 1.3 $M_\odot$ ($\alpha_\mathrm{IGIMF}$) for the
  SFR values of Tab.~\ref{tab:first} and some observational
  constrains in dependence of the total SFR. The {\it shaded region}
  between the {\it solid lines} marks the range of model results for
  an ECMF slope $\beta$ = 2.0 while the {\it dotted lines} show the
  full envelope constrained by all models. The three {\it asterisks}
  are from \citet{Da07}, the one {\it triangle} around $\alpha_3
  \approx 0.7$ is from \citet{VD07}, the {\it filled circle} with
  error bars is from \citet{BKM07} while the {\it open circle} with
  error bars corresponds to the \citet{BLF05} result. The {\it dashed}
  horizontal line marks the \citet{WHT08} constrain with the light
  {\it dashed lines} 0.15 dex above and below being their uncertainty
  range.}
\label{fig:result}
\end{center}
\end{figure}

With this contribution the SFR dependent IGIMF theory is extended into
the starburst regime. The SFR dependent IGIMF, which produces results
agreeing with observations for massive and dwarf galaxies
\citep{KW03,WK05a,KWK05,RCK09}, leads to top-heavy 
galaxy-wide IMFs in massively starbursting galaxies when combined with 
the cluster-mass dependent high-mass IMF slope for globular clusters
and UCDs (eq.~\ref{eq:a3M}) by Marks, Kroupa \& Dabringhausen
(in preparation) and a SFR dependent lower
limit of the ECMF or flat slopes of the ECMF. This is interesting because
some observational results
\citep{FBB03,ATB04,TBM04,Lo06,SMB05,NS05,ABA07} indeed suggest the 
IGIMF to be top-heavy when SFR $\simgreat$ 100
$M_\odot~yr^{-1}$. Current observational results from
cosmological studies are very difficult to compare with the models
because the observations do not measure IMF slopes and SFRs for
individual galaxies but study indirect evidence (luminosities,
chemical enrichment, etc.) for whole populations and average the
results over the galaxy luminosity function. The most-massive star
cluster for which the IMF has been determined by star counts, R136 in
the LMC, shows no evidence for a non-Salpeter slope of the high-mass
stars \citep{SMB99}. However, this cluster falls below the mass
  limit above which the parametrisation  
chosen here (eq.~\ref{eq:a3M}) implies a top-heavy IMF.
Additionally, no common 
description is used in the literature to characterize the
top-heaviness. Some authors vary the peak of the IMF while others vary
the slope or the lower and upper mass limit. Some available data points 
are shown in Fig.~\ref{fig:result} together with the model
predictions. The cosmologically interpreted observations by
\citet[][{\it triangle}]{VD07} and \citet[][{\it open circle}]{BLF05}
suggest rather flat IMFs already for relatively low SFRs and are
difficult to reconcile with the here presented model even for values
of $M_\mathrm{ecl,min}$ which are very high ($10^5$ $M_\odot$). But as
can also be seen in Fig.~\ref{fig:result} the IGIMF results are in good
agreement with the constraints for the Galactic and M31 bulge \citep[{\it
filled circle} with error-bar,][]{BKM07}, as well as the \citet{WHT08}
constrains for the present-day mass density from the cosmological SFH
({\it dashed lines}). The GAMA-team finds a very similar trend of
  decreasing slope with SFR as the models in their sample of $\sim$
  40000 galaxies \citep{GHS10}. Likewise, the \citet[][{\it asterisks}]{Da07}
study of the amplitude of the galaxy stellar mass-star formation rate
relationship seems to be in reasonable agreement with the
model. Though for this agreement it is necessary that for galaxies
with SFRs above 10 $M_\odot~yr^{-1}$ the ECMF has either a lower limit
larger then 10000 $M_\odot$ or a slope flatter then 1.5. Generally, it
should be possible to test these scenarios observationally by
observing the ECMF.

Summarising, the main physical reason why the IGIMF becomes
top-heavy at galaxy-wide SFRs $>$ 10 $M_\odot$ yr$^{-1}$ is the
formation of very massive star clusters with masses M$_\mathrm{ecl}$
$> 10^6$ $M_\odot$. To achieve such high SFRs a galaxy with neutral gas needs
to either globally be unstable (e.g.~at early cosmological times when
the gas fraction was very high) or be compressed globally due to an
external tidal force field. Under these conditions the pressurised
interstellar medium will collapse to massively giant molecular cloud
complexes with masses of 10$^8$ $M_\odot$ or larger. In these complexes,
ultra-compact-dwarf galaxy type star clusters may form with individual
masses $>10^6$ $M_\odot$, and these would be having top-heavy stellar
IMFs \citep{DKB09}, as introduced into the IGIMF theory here. 
It may also be possible that at the same time the formed star-cluster
mass function becomes bottom light under extreme SFRs, which may be
due to the conditions for forming low-mass ($<$100-1000
$M_\odot$) clusters not being available, perhaps due to the intense stellar
feedback which may suppress the formation of such low-mass molecular
cloud cores.

It should be noted here that the two rather extreme top-heavy IMF
suggestions \citep{BLF05,VD07} with $\alpha_\mathrm{empirical} \le$ 1
are difficult to reconcile with the existence of globular clusters
today. To obtain such flat IGIMFs the massive star clusters 
would need to have such flat IMFs that more than 90\% of their
mass would be in stars which explode as supernovae. Such extreme
mass-loss would quickly fully disperse these clusters, leaving no
globular clusters behind.

\section*{Acknowledgements}
This work was funded by the Chilean FONDECYT grand 3060096 and the
European Commission Marie Curie Research Training Grant CONSTELLATION 
(MRTN-CT-2006-035890).

\begin{appendix}
\section{The canonical IMF}
\label{app:IMF}
The following multi-component power-law IMF is used throughout the paper:

{\small
\begin{equation}
\xi(m) = k \left\{\begin{array}{ll}
k^{'}\left(\frac{m}{m_{\rm H}} \right)^{-\alpha_{0}}&\hspace{-0.25cm},m_{\rm
  low} \le m < m_{\rm H},\\
\left(\frac{m}{m_{\rm H}} \right)^{-\alpha_{1}}&\hspace{-0.25cm},m_{\rm
  H} \le m < m_{0},\\
\left(\frac{m_{0}}{m_{\rm H}} \right)^{-\alpha_{1}}
  \left(\frac{m}{m_{0}} \right)^{-\alpha_{2}}&\hspace{-0.25cm},m_{0}
  \le m < m_{1},\\ 
\left(\frac{m_{0}}{m_{\rm H}} \right)^{-\alpha_{1}}
    \left(\frac{m_{1}}{m_{0}} \right)^{-\alpha_{2}}
    \left(\frac{m}{m_{1}} \right)^{-\alpha_{3}}&\hspace{-0.25cm},m_{1}
    \le m < m_{\rm max},\\ 
\end{array} \right. 
\label{eq:4pow}
\end{equation}
\noindent with exponents
\begin{equation}
          \begin{array}{l@{\quad\quad,\quad}l}
\alpha_0 = +0.30&0.01 \le m/{M}_\odot < 0.08,\\
\alpha_1 = +1.30&0.08 \le m/{M}_\odot < 0.50,\\
\alpha_2 = +2.35&0.50 \le m/{M}_\odot < 1.00,\\
\alpha_3 = +2.35&1.00 \le m/{M}_\odot \le m_\mathrm{max}.\\
          \end{array}
\label{eq:imf}
\end{equation}}
\noindent where $dN = \xi(m)\,dm$ is the number of stars in the mass
interval $m$ to $m + dm$. The exponents $\alpha_{\rm i}$ represent the
standard or canonical IMF \citep{Kr01,Kr02}. Though, $\alpha_3$ is
kept constant at 2.35 only for star clusters with $M_\mathrm{ecl}$
less than 2$\times 10^5$ $M_\odot$. For more massive clusters
$\alpha_3$ is changed with cluster mass according to eq.~\ref{eq:a3M}.

The advantage of such a multi-part power-law description are the easy
integrability and, more importantly, that {\it different parts of the
  IMF can be changed readily without affecting other parts}. Note that
this form is a two-part power-law in the stellar regime, and that
brown dwarfs contribute about 4 per cent by mass only and that a
discontinuity near $m_\mathrm{H}$ implies brown dwarfs to be a
separate population \citep[$k^{'} \sim \frac{1}{3}$,][]{TK07,TK08}. A
log-normal form in the stellar regime but below 1~$M_{\odot}$ with a
power-law extension to high masses was suggested by \citet{Ch03} but
is virtually identical to the canonical IMF \citep[fig.~8 in ][]{DHK08}.

The observed IMF is today understood to be an invariant
Salpeter/Massey power-law slope \citep{Sal55,Mass03} above
$0.5\,M_\odot$, being independent of the cluster density and
metallicity for metallicities $Z \simgreat 0.002$
\citep{MH98,SND00,SND02,PaZa01,Mass98,Mass02,Mass03,WGH02,BMK03,PBK04,PAK06}. 
Furthermore, un-resolved multiple stars in the young star clusters are
not able to mask a significantly different slope for massive stars
\citep{MA08,WK07c}. \citet{Kr02} has shown that there are no trends
with physical conditions and that measured high-mass slopes,
$\alpha_3$, are a Gaussian distribution about the Salpeter value thus
allowing us to assume for now that the stellar IMF is invariant and
universal in each cluster. There is evidence of a maximal mass for
stars \citep[$m_{\rm max*}\,\approx\,150\,M_{\odot}$,][]{WK04}, a
result later confirmed by several independent studies
\citep{OC05,Fi05,Ko06}. However, according to \citet{CSH10}
$m_\mathrm{max*}$ may also be as high as 300 $M_\odot$.
\end{appendix}

\bibliography{mybiblio}

\bsp
\label{lastpage}
\end{document}